\documentclass[letterpaper, 10pt, conference]{ieeeconf}  
\IEEEoverridecommandlockouts
\usepackage{cite}
\usepackage{graphicx}
\usepackage{array}
\usepackage{hyperref}
\usepackage{arydshln}
\usepackage{float}
\usepackage{blindtext}
\usepackage{hyphenat}
\usepackage{xspace}

\let\labelindent\relax

\usepackage{tikz}
\usepackage{xspace}
\usepackage{xcolor}
\usepackage{amsthm}
\usepackage{amsmath,amssymb,amsfonts}
\usepackage{mathrsfs,mathtools}
\usepackage{mathalfa}
\usepackage{euscript}
\usepackage{csquotes}
\usepackage{enumitem}
\usepackage[normalem]{ulem}

\DeclareRobustCommand{\legendline}[1]{\hspace{-0pt}\tikz[#1,line width=0.4mm,baseline=-0.5ex]{\draw (0,0) -- (.2,0);}\hspace{-0pt}}

\definecolor{mblue}{rgb}{0,0.4470,0.7410}
\definecolor{morange}{rgb}{0.8500,0.3250,0.0980}
\definecolor{myellow}{rgb}{0.9290,0.6940,0.1250}
\definecolor{mpurple}{rgb}{0.4940,0.1840,0.5560}
\definecolor{mgreen}{rgb}{0.4660,0.6740,0.1880}
\definecolor{mcyan}{rgb}{0.3010,0.7450,0.9330}
\definecolor{mred}{rgb}{0.6350,0.0780,0.1840}
\definecolor{mgreenblue}{rgb}{0.0,1.0,0.5}
\definecolor{parulablue}{rgb}{0.2431,0.1490,0.6588}
\definecolor{parulalblue}{RGB}{39,151,235}
\definecolor{parulagreen}{RGB}{129,204,89}
\definecolor{parulayellow}{RGB}{249,251,21}

\definecolor{cblue}{rgb}{0,0.9,1}
\definecolor{corange}{rgb}{1,0.7,0}
\definecolor{cgray}{rgb}{0.5,0.5,0.5}

\theoremstyle{definition}
\newtheorem{definition}{Definition}

\theoremstyle{plain}
\newtheorem{theorem}{Theorem}

\newtheorem{conjecture}{Conjecture}
\newtheorem{corollary}{Corollary}

\newtheorem{assumption}{Assumption}
\theoremstyle{remark}
\newtheorem{remark}{Remark}

\newcommand{\comment}[1]{}

\newcounter{ass}

\newcommand{\mc}[1]{\mathcal{#1}}
\newcommand{\mf}[1]{\mathfrak{#1}}
\newcommand{\mr}[1]{\mathrm{#1}}
\newcommand{\mb}[1]{\mathbb{#1}}
\newcommand{\ms}[1]{\mathscr{#1}}

\newcommand{\mt}[1]{\mathtt{#1}}
\newcommand{\mbf}[1]{\mathbf{#1}}
\newcommand{\meu}[1]{\EuScript{#1}}
\DeclareFontFamily{U}{txcal}{\skewchar \font =45}
\DeclareFontShape{U}{txcal}{m}{n}{<-> txr-cal}{}
\DeclareMathAlphabet{\mathcalpxtx}{U}{txcal}{m}{n}

\newcommand{\dif}{\mr{d}}

\newcommand{\tPartial}[3]{\tfrac{\partial^{#1} #2}{\partial #3^{#1}}}
\newcommand{\posdef}{\succ}

\newcommand{\unaryminus}{\scalebox{0.65}[1]{\ensuremath{\,-}}}

\newcommand{\diag}{\mr{diag}}

\newcommand{\col}{\mr{col}}

\DeclareMathOperator{\sind}{sind}

\newcommand{\kron}{\otimes} %

\newcommand{\dnx}{n_\mr{x}}
\newcommand{\dny}{n_\mr{y}}
\newcommand{\dnu}{n_\mr{u}}
\newcommand{\dnp}{n_\mr{p}}

\newcommand{\Z}{\mb{Z}}

\newcommand{\R}{\mb{R}}

\newcommand{\Up}{U^{\mt{p}}}
\newcommand{\Xp}{X^{\mt{p}}}
\newcommand{\Xsdelta}{\overrightarrow{X}_{\!\Delta}}
\newcommand{\deltax}{\Delta x}
\newcommand{\deltau}{\Delta u}
\newcommand{\deltay}{\Delta y}

\newcommand{\mG}{\mc{G}_\Delta}
\newcommand{\nldataset}[1]{\mc{D}_{#1}^{\textsc{nl}}}
\newcommand{\ddataset}[1]{\mc{D}_{#1}^{\Delta}}

\newcommand{\deflen}[2]{%
    \expandafter\newlength\csname #1\endcsname
    \expandafter\setlength\csname #1\endcsname{#2}%
}

\newcommand{\sumop}{\mbf{\Sigma}}
\newcommand{\difop}{\mbf{\Delta}}

\ifx\usecoloredremarks\undefined
    
    \newcommand{\chris}[1]{#1}

    \newcommand{\new}[1]{#1}
\else
    \newcommand{\chris}[1]{{\color{mblue} #1}}

    \newcommand{\new}[1]{{\color{blue} #1}}
\fi

\let\OLDthebibliography\thebibliography
\renewcommand\thebibliography[1]{
  \vspace{-0.5mm}\OLDthebibliography{#1}\vspace{-0.7mm}
  \setlength{\parskip}{0pt}
  \setlength{\itemsep}{0pt plus 0.1ex}
}

\title{Direct data-driven state-feedback control of general nonlinear systems \vspace{-3mm}}

\author{%
    Chris Verhoek, Patrick J. W. Koelewijn, Sofie Haesaert, and Roland T\'oth%
    \thanks{This work was supported by  the European Research Council (ERC) under the European Union's Horizon 2020 research and innovation programme (grant agreement nr. 714663) and the European Union within the framework of the National Laboratory for Autonomous Systems (RRF-2.3.1-21-2022-00002). %
    The authors are with the Control Systems group in the Dept. of Electrical Engineering at the Eindhoven University of Technology, The Netherlands. Roland T\'oth is also with the Institute for Computer Science and Control, Budapest, Hungary.}%
    \thanks{Corresponding author: C. Verhoek (\texttt{c.verhoek@tue.nl}).}%
}

\begin{document}
\maketitle
\thispagestyle{empty}
\pagestyle{empty}

\begin{abstract}
    Through the use of the Fundamental Lemma for linear systems, a direct data-driven state-feedback control synthesis method is presented for a rather general class of nonlinear (NL) systems. The core idea is to develop a data-driven representation of the so-called velocity-form, i.e., the time-difference dynamics, of the NL system, which is shown to admit a direct \emph{linear parameter-varying} (LPV) representation. By applying the LPV extension of the Fundamental Lemma in this velocity domain, a state-feedback controller is directly synthesized to provide asymptotic stability and dissipativity of the velocity-form. By using realization theory, the synthesized controller is realized as a NL state-feedback law for the  original unknown NL system with guarantees of \emph{universal shifted stability and dissipativity}, i.e., stability and dissipativity w.r.t. any (forced) equilibrium point, of the closed-loop behavior. This is achieved by the use of a single sequence of data from the system and a predefined basis function set to span the scheduling map. The applicability of the results is demonstrated on a simulation example of an unbalanced disc.

\end{abstract}

\deflen{rmwhitebfsec}{-0.5mm} %
\deflen{rmwhiteafsec}{-0.7mm} %
\deflen{rmwhitebfssec}{-0.5mm} %
\deflen{rmwhiteafssec}{-0.7mm} %

\vspace{\rmwhitebfsec}\section{Introduction}\label{sec:introduction}\vspace{\rmwhiteafsec}
Due to the ever-increasing performance requirements, control problems in engineering are getting increasingly more complex, with the need to precisely address \emph{nonlinear}~(NL) aspects of the behavior of the underlying systems. This in turn also 
requires %
accurate modeling of such NL behaviors, which often becomes cumbersome or even impossible with first-principle modeling techniques. While data-driven methods provide an alternative, in the absence of a mature NL identification for control theory, it is often  difficult to decide which part of the behavior is crucial to be captured for control design and how the uncertainty of the estimated model influences the subsequent control synthesis.
For this reason, data-driven control methods have been developed to design controllers \emph{directly} from data, eliminating the need of a modeling step. In the \emph{linear time-invariant} (LTI) case, the \emph{Fundamental Lemma}~\cite{WillemsRapisardaMarkovskyMoor2005} has proven to be a key result, allowing for direct data-driven analysis and control synthesis with stability and performance guarantees, see~\cite{MarkovskyDorfler2021Review}. %
Besides of promising approaches based on feedback and online linearizations, or polynomial bases~\cite{de2023learning, berberich2022linear, markovsky2021data}, an analogous result for general NL systems has not been achieved yet.

In this paper, we propose a novel extension of the Fundamental Lemma to a wide class of  \emph{discrete-time} (DT) NL systems that can be described in a state-space form with differentiable state transition and output functions. Our result is based on the use of  
the velocity-form of the NL system, which describes the time-difference
dynamics of the system and it has two important properties: %
\textit{(i)}~stability and performance of the velocity-form imply \emph{universal shifted}, i.e., equilibrium-independent,  stability and performance of the original NL system~\cite{Koelewijn2023,Koelewijn2023a,Simpson-Porco2019}, %
\textit{(ii)}~the velocity-form naturally results in a \emph{linear parameter-varying} (LPV) system. By calculating time-differences of the data from the underlying NL system, which characterizes the velocity form, our first contribution (C1) is to show that the resulting data-equations allow for convex data-driven analysis and controller synthesis by the use of the recently introduced LPV Fundamental Lemma \cite{VerhoekTothHaesaertKoch2021} due to property \emph{(i)}. Then, by exploiting \emph{(ii)}, our second main contribution (C2) is to show that the data-driven controller for the velocity-form, obtained in the previous step, exhibits a computable realization, and to prove that this realization provides {universal shifted} guarantees for  closed-loop control of the original NL system.

In Section~\ref{sec:problemstatement}{,} {we} formalize the NL data-driven control problem that we intend to solve. Section~\ref{sec:preliminaries} introduces the data-based representation of the velocity-form of the NL system using {an} LPV embedding and the {LPV} Fundamental Lemma. Section~\ref{sec:mainresults} uses the data-driven representation to synthesize a state-feedback controller for the velocity-form, which by realization to a NL state-feedback law provides {equilibrium independent} guarantees. Section~\ref{sec:examples} demonstrates the applicability of the results in a simulation example based on an unbalanced disc system, {while} the conclusions on the achieved results are given in Section~\ref{sec:conclusion}. 

\subsubsection*{Notation}
The set of integers is denoted by $\Z$, while the set of real numbers is denoted by $\R$. Moreover, $\R_0^+=[0,\infty)\subset\R$. A function $f:\R^{p}\to\R^q$ is in $\mc{C}^n$ if it is $n$-times continuously differentiable, while $f:\R^p\to\R$ belongs to the class $\mc{Q}_{x_\ast}$ if it is positive definite and decrescent w.r.t. $x_\ast\in\R^p$ (see \cite{SchererWeiland2021}). \new{$\col(x_1, \dots, x_n)$ denotes $[x_1^\top \hspace{0.2mm} \cdots\hspace{0.7mm} x_n^\top]^\top$.}
\vspace{\rmwhitebfsec}\section{Problem statement}\label{sec:problemstatement}\vspace{\rmwhiteafsec}
Consider a DT NL system\footnote{As we intent to establish the core concepts on data-driven control of NL systems {via the Fundamental Lemma}, in this work, we do not consider disturbance or noise signals in \eqref{eq:NL_sys}. The extensions towards noise-affected systems are objective of future {research}, e.g., based on \cite{guo2021data}.}, defined in terms of the state-space representation
\begin{equation}\label{eq:NL_sys}
    x_{k+1}  = f(x_k,u_k), \quad   y_k  = x_k, 
\end{equation}
where $x_k\in\mb{X}\subseteq \mb{R}^{n_\mr{x}}$ is the state, $u_k\in\mb{U} \subseteq \mb{R}^{n_\mr{u}}$ is the input and $y_k\in\mb{Y}=\mb{X}$ is the observed output at time moment $k\in\mathbb{Z}$. Here, $y_k$ is assumed to provide full state observation. $\mb{X}$ and $\mb{U}$ are considered to be open sets containing the origin, and $f: \mb{X} \times \mb{U} \to \mb{X}$ is assumed to be a $\mc{C}^1$ function. 
The \emph{behavior}, i.e., the set of all solution trajectories of \eqref{eq:NL_sys}, is %
\begin{equation}
    \hspace{-1mm}\mf{B}=\left\{ (x,u, y)\in ( \mb{X}\times\mb{U}\times\mb{Y})^{\mathbb{Z}}\mid\text{\eqref{eq:NL_sys} holds } \forall k\in\mathbb{Z}  \right\}.
\end{equation}
The set of all (forced) equilibrium points of \eqref{eq:NL_sys} \new{is given by}
\begin{equation*}
    \ms{E} = \{ (x_\ast,u_\ast, y_\ast)\in \mb{X}\times\mb{U}\times\mb{Y} \mid x_\ast  = f(x_\ast,u_\ast), \, y_\ast=x_\ast \}.
\end{equation*}
Furthermore, let $\mb{X}^\ast=\pi_{x_\ast}\ms{E}$, $\mb{U}^\ast=\pi_{u_\ast}\ms{E}$, $\mb{Y}^\ast=\pi_{y_\ast}\ms{E}$, \new{where $\pi$ is the projection operator w.r.t.  specific variables.}

As highlighted in Section~\ref{sec:introduction}, analyzing the time-difference dynamics of \eqref{eq:NL_sys} allows for giving equilibrium independent guarantees on \eqref{eq:NL_sys}~\cite{Koelewijn2023,Koelewijn2023a}. {For this purpose,} we introduce the {so-called} \emph{velocity-form} of \eqref{eq:NL_sys} {that will be an important ingredient in our proposed method.}
For the increments
\begin{equation}\label{eq:deltasignals}
    \deltau_k = u_{k}-u_{k-1}, \ \deltax_k= x_{k}-x_{k-1}, \ \deltay_k= y_{k}-y_{k-1},
\end{equation}
we obtain the time-difference dynamics as
\begin{equation}\label{eq:NL_sys_diff}
    \deltax_{k+1}  = f(x_{k},u_{k})-f(x_{k-1},u_{k-1}), \quad \deltay_k  = \deltax_k.
\end{equation}
By the use of the \emph{Fundamental Theorem of Calculus}, e.g., {see} \cite[Lem.~C.1.1]{Koelewijn2023}, \eqref{eq:NL_sys_diff} can be rewritten in the equivalent \emph{velocity-form}:
\begin{subequations}\label{eq:NL_sys_velocity}
\begin{align}
    \deltax_{k+1} & = A_\mr{v}(\xi_{k}, \xi_{k-1})\deltax_k + B_\mr{v}(\xi_{k}, \xi_{k-1})\deltau_k, \label{eq:NL_sys_velocity:state}\\ 
    \deltay_k & = \deltax_k,
\end{align}
where $\xi_k=\col(x_k, u_k)$, and
\begin{align}
    A_\mr{v}(x_{k}, u_{k}, x_{k-1}, u_{k-1}) &= \int_0^1\tPartial{}{f}{x}\big(\bar{x}_k(\lambda), \bar{u}_k(\lambda)\big)\dif \lambda, \\
    B_\mr{v}(x_{k}, u_{k}, x_{k-1}, u_{k-1}) &= \int_0^1\tPartial{}{f}{u}\big(\bar{x}_k(\lambda), \bar{u}_k(\lambda)\big)\dif \lambda,
\end{align}
with $\bar{x}_k(\lambda)=x_{k-1} + \lambda(x_{k}-x_{k-1})$ and $\bar{u}_k(\lambda)=u_{k-1} + \lambda(u_{k}-u_{k-1})$, $\lambda\in [0,1]$.
\end{subequations}
The solutions of \eqref{eq:NL_sys_velocity} are collected in the \emph{velocity behavior} $\mf{B}_\Delta$, which is defined as
\begin{multline}
    \hspace{-1mm}\mf{B}_\Delta=\big\{ (\deltax,\deltau,\deltay)\in (\R^{\dnx}\times\R^{\dnu}\times\R^{\dny})^{\mathbb{Z}} \mid \text{the}\hspace{1mm} \\ \text{relations in \eqref{eq:deltasignals} hold }\forall k\in\mathbb{Z}, (x,u,y)\in\mf{B} \big\}.\hspace{-1mm}
\end{multline}
Analyzing stability and performance of the velocity-form \eqref{eq:NL_sys_velocity} by means of the concept of dissipativity yields universal guarantees on \eqref{eq:NL_sys}. Hence, consider the following definitions:
\begin{definition}\label{def:velocity_stab}
    The system \eqref{eq:NL_sys} is velocity-stable, if 
    \eqref{eq:NL_sys_velocity} is stable with $\deltau=0$, i.e., for each $\epsilon>0$ there exists a $\delta(\epsilon)$ such that $\|\deltax_{k_0}\|\le\delta(\epsilon) \Rightarrow \|\deltax_{k}\|\le\epsilon,\, \forall k\ge k_0$. It is asymptotic velocity-stable, if it is velocity-stable and for $\deltau=0$ we have $\lim_{k\to\infty}\|\deltax_{k}\|=0$.
\end{definition}
{\begin{definition}\label{def:velocity_diss}
    The system \eqref{eq:NL_sys} is velocity-dissipative w.r.t. the supply function $\meu{S}_\mr{v}:\mb{R}^{\dnu}\times\mb{R}^{\dny}\to\mb{R}$, if there exists a storage function $\meu{V}_\mr{v}:\R^{\dnx}\to\mb{R}_0^+$ with $\meu{V}_\mr{v}\in\mc{C}_0$, $\meu{V}_\mr{v}\in\mc{Q}_{0}$, such that 
    \begin{equation}\label{eq:VD}
        \meu{V}_\mr{v}(\deltax_{k_1+1}) - \meu{V}_\mr{v}(\deltax_{k_0}) {\ \leq \ }{\textstyle\sum_{k=k_0}^{k_1}}\meu{S}_\mr{v}(\deltau_k, \deltay_k),
    \end{equation}
    for all $k_0, k_1\in\mb{Z}$, $k_0\le k_1$ and $(\deltax,\deltau,\deltay)\in\mf{B}_\Delta$. %
\end{definition}}
{It is well-known that dissipativity implies asymptotic stability if $\meu{S}_\mr{v}$ is a negative definite function under %
zero input, i.e., there is a strictly decreasing $\alpha:\mathbb{R}_0^+\rightarrow \mathbb{R}_0^+$ with $\alpha(0)=0$, s.t. $\meu{S}_\mr{v}(0, \deltay)<\alpha(\|\deltay\|)$ for all $\deltay\in\R^{\dny}$ %
\cite{Koelewijn2023}.

It has been shown in \cite{Koelewijn2023} that velocity-stability and velocity-dissipativity implies strong \new{equilibrium independent stability and performance} notions in terms of
\emph{universal shifted (asymptotic) stability} (US(A)S) and \emph{universal shifted dissipativity} (USD), which are defined as:
\begin{definition}\label{def:USS}
    The system \eqref{eq:NL_sys} is USS if it is stable w.r.t. all $(x_\ast,u_\ast, y_\ast)\in\ms{E}$, i.e., if for each $\epsilon>0$ there exists a $\delta(\epsilon)$ such that $\|x_{k_0}-x_\ast\|\le\delta(\epsilon) \Rightarrow \|x_{k}-x_\ast\|\le\epsilon$, $\forall k\ge k_0$. It is USAS if it is USS and for all $(x_\ast,u_\ast, y_\ast)\in\ms{E}$ we have $\lim_{k\to\infty}\|x_{k}-x_\ast\|=0$ with $(x,u,y)\in\mf{B}$ for which $u\equiv u_\ast$. %
\end{definition}
\begin{definition}\label{def:USD}
    The system \eqref{eq:NL_sys} is USD w.r.t. the supply function $\meu{S}_\mr{s}:\mb{U}\times\mb{U}^*\times\mb{Y}\times\mb{Y}^\ast\to\mb{R}$, if there exists a storage function $\meu{V}_\mr{s}:\mb{X}\times\mb{U}^*\to\mb{R}_0^+$, which $\forall (x_\ast,u_\ast)\in \pi_{x_\ast,u_\ast}\ms{E}$ satisfies
 $\meu{V}_\mr{s}(\cdot,u_\ast)\in\mc{C}_0$, $\meu{V}_\mr{s}(\cdot,u_\ast)\in\mc{Q}_{x_\ast}$, and %
    \begin{equation}\label{eq:USD}
        \hspace{-1.5mm}\meu{V}_\mr{s}({x_{k_1+1}}, u_\ast) \unaryminus \meu{V}_\mr{s}(x_{k_0}, u_\ast)  \leq {\textstyle\sum_{k=k_0}^{k_1}}\meu{S}_\mr{s}(u_k, u_\ast, y_k, y_\ast),\hspace{-.5mm}
    \end{equation}
    for all $k_0, k_1\in\mb{Z}$, $k_0\le k_1$ and $(x,u,y)\in\mf{B}$.
\end{definition}
The key-observation is that in \cite{Koelewijn2023} it is proven that velocity-dissipativity \emph{implies} US(A)S, i.e., (asymptotic) stability of \eqref{eq:NL_sys} w.r.t. any equilibrium point in $\ms{E}$. Furthermore, under certain conditions%
\footnote{See \cite[Sec.~8.3]{Koelewijn2023} for the conditions, and the discussion in Section~\ref{ss:stabandperf}.}
velocity-dissipativity implies USD, i.e., performance of~\eqref{eq:NL_sys} w.r.t. any equilibrium point in $\ms{E}$. This allows for the design and synthesis of  controllers for \eqref{eq:NL_sys} through the velocity-form, \new{which, after appropriate realization, will} %
guarantee \new{universal shifted} stability and performance of the {closed-loop} NL system. In \cite{Koelewijn2023}, this has been {accomplished} in the model-based setting using {an} LPV  form of \eqref{eq:NL_sys_velocity}. {We aim to extend this result} to the data-based setting by solving the following problem:

\subsubsection*{Problem statement} 
Consider a system represented by \eqref{eq:NL_sys} {from which} $N$ samples of input-state data {have been obtained and} collected in the \emph{data-dictionary} $\nldataset{N}= \{u^\mr{d}_k, x^\mr{d}_k\}_{k=1}^N$. How to synthesize a state-feedback controller for \eqref{eq:NL_sys}, {purely} based on $\nldataset{N}$, such that the controller guarantees {universal shifted} stability and performance of the closed-loop system?

\vspace{\rmwhitebfsec}\section{Data-based velocity representations}\label{sec:preliminaries}\vspace{\rmwhiteafsec}
To realize our objective, we first show that the velocity-form admits an LPV embedding and that we can obtain an LPV
data-driven representation of %
$\mf{B}_\Delta$ purely based on $\nldataset{N}$.

\vspace{\rmwhitebfssec}\subsection{LPV embedding of the velocity-form}\vspace{\rmwhiteafssec}
To apply the embedding principle, we will need to start with the following assumption:
\begin{assumption}\label{ass:schedulingmap}
    We are given a set of basis functions $\psi_1, \dots, \psi_{\dnp}$ with $\psi: (\mb{X}  \times \mb{U})^2\to\R^{\dnp}$, such that there exist  $A_0, \dots, A_{\dnp}\in\R^{\dnx\times\dnx}$ and $B_0, \dots, B_{\dnp}\in\R^{\dnx\times\dnu}$ for which
    \begin{subequations}\label{eq:decompab}
    \begin{align}
        A_\mr{v}(\xi_{k}, \xi_{k-1}) & = A_0 + {\textstyle\sum_{i=1}^{\dnp}}A_i\psi_i(\xi_{k}, \xi_{k-1}), \\
        B_\mr{v}(\xi_{k}, \xi_{k-1}) & = B_0 + {\textstyle\sum_{i=1}^{\dnp}}B_i\psi_i(\xi_{k}, \xi_{k-1}),
    \end{align}
    \end{subequations}
    with $\xi_k=\col(x_k, u_k)$. 
\end{assumption}
\begin{remark}
    {By} %
    a polynomial basis {set} in $x_k,u_k,x_{k-1},u_{k-1}$, one can \emph{approximate} any $A_\mr{v}$ and $B_\mr{v}$ in the velocity form~\eqref{eq:NL_sys_velocity} under the condition that $f\!\in\!\mc{C}^\infty$, which can be easily shown based on the Taylor series of $f$. Alternatively, one can use kernel-based methods to learn $\psi$ from data.
    Hence, explicit prior knowledge of $f$ is not {necessary} for {choosing an effective} %
   $\psi$. Only the number of basis $\dnp$, governing the approximation error, is required to be determined in advance.%
\end{remark}
Based on Assumption~\ref{ass:schedulingmap}, we use the given set of functions to define a so-called \emph{scheduling variable}, a signal that can represent all the variation of the nonlinearities in \eqref{eq:decompab}: %
\begin{equation}\label{eq:computation_p}
   p_k=\psi(x_k, u_{k}, x_{k-1}, u_{k-1})\in\mb{P} \subseteq \R^{\dnp}. 
\end{equation}
Note that $p_k$ can be computed from measurements of $y_k=x_k$ and $u_k$ through $\psi$, hence using %
the available data set $\nldataset{N}$. Here, $\mb{P}$ can be constructed as the convex hull of the image of $\mb{X} \times \mb{U}$ or $\nldataset{N}$ through $\psi$, where convexity is required by the analysis and synthesis tools we will use in Section~\ref{sec:mainresults}. %

 With \eqref{eq:computation_p}, the LPV embedding of \eqref{eq:NL_sys_velocity} is {formulated as}
\begin{subequations}\label{eq:NL_sys_LPV}
\begin{align}
    \deltax_{k+1} & = \bar{A}_\mr{v}(p_k)\deltax_k + \bar{B}_\mr{v}(p_k)\deltau_k, \label{eq:NL_sys_LPV:state}\\ 
    \deltay_k & = \deltax_k,
\end{align}
\end{subequations}
with $\bar{A}_\mr{v}(p_k)  = A_0+\sum_{i=1}^{\dnp}A_i p_{i,k}$ and $\bar{B}_\mr{v}(p_k)  = B_0+\sum_{i=1}^{\dnp}B_i p_{i,k}$. {To make \eqref{eq:NL_sys_LPV}  a linear surrogate representation of \eqref{eq:NL_sys_diff}, in the LPV framework, $p_k\in\mb{P}$ in \eqref{eq:NL_sys_LPV} is assumed to vary independently from $(\Delta x_k, \Delta u_k)$.} The resulting behavior of \eqref{eq:NL_sys_LPV} is defined as
\begin{multline*}
    \mf{B}_{\textsc{lpv}}\!=\! \{(\deltax,\deltau,\deltay)\in(\mb{X}\times\mb{U}\times\mb{Y})^\mb{Z} \mid \exists p \in \mb{P}^\mb{Z} \text{ s.t.}  \\\text{\eqref{eq:NL_sys_LPV} holds } \forall k\in\mathbb{Z}\}.
\end{multline*}
{The} assumption {of} the independent variation of $p_k$ {implies that} $\mf{B}_\Delta\subset\mf{B}_{\textsc{lpv}}$, {resulting in an embedding of the %
velocity behavior $\mf{B}_\Delta$ into a solution set of a linear representation. {While} the price for this linearity is payed in the conservatism %
of the resulting LPV representation,} %
linearity {in itself enables the derivation of} %
{a} data-driven representation concept {through the LPV extension of the Fundamental Lemma}. 
 \begin{remark}{The velocity-form is key to accomplish the LPV embedding, because \textit{(i)} \eqref{eq:NL_sys_velocity:state} naturally appears in an LPV form compared to the required non-unique factorization of $f$ and $h$ for the \emph{direct} LPV embedding of \eqref{eq:NL_sys} (as is used in, e.g., \cite{Verhoek2022_DDLPVstatefb_experiment}), and \textit{(ii)} ensuring (asymptotic) stability and dissipativity guarantees on \eqref{eq:NL_sys_LPV} results in \new{equilibrium independent} guarantees on \eqref{eq:NL_sys}, while this is not the case with a direct LPV embedding and LPV analysis of \eqref{eq:NL_sys}, see~\cite{Koelewijn2020_pitfalls} for further details.}
 \end{remark}

\vspace{\rmwhitebfssec}\subsection{Data-driven \new{closed-loop} velocity representations}\vspace{\rmwhiteafssec}

To make a data-driven synthesis for the velocity form and a subsequent realization of the controller for the original NL system \eqref{eq:NL_sys} possible, we require as a first step a data-driven representation of \eqref{eq:NL_sys_diff} in closed-loop with the to-be-designed controller. By exploiting the LPV embedding concept  \eqref{eq:NL_sys_LPV} of \eqref{eq:NL_sys}, we can derive such a closed-loop representation  based on \cite{Verhoek2022_DDLPVstatefb}
using $\nldataset{N+1}=\{u^\mr{d}_k, x^\mr{d}_k\}_{k=1}^{N+1}$, measured from %
\eqref{eq:NL_sys}. %

Based on $\mc{D}_{N+1}^{\textsc{nl}}$, we can construct the signals that constitute \eqref{eq:NL_sys_LPV}, %
{resulting in the} %
data-dictionary $\mc{D}_N^\Delta = \{\deltax^\mr{d}_k, p^\mr{d}_k, \deltau^\mr{d}_k \}_{k=2}^{N+1}$ {and} %
the data matrices
\begin{subequations}\label{eq:datamatrices}\deflen{savespaceeqeight}{-1.5mm}%
\begin{align}
	\hspace{\savespaceeqeight}U_\Delta   & = \begin{bmatrix} \deltau^\mr{d}_2 & \cdots  & \deltau^\mr{d}_{N} \end{bmatrix}\in\R^{\dnu\times N-1}, \\
	\hspace{\savespaceeqeight}\Up_\Delta & = \begin{bmatrix} p^\mr{d}_2\otimes \deltau^\mr{d}_2 & \cdots &  p^\mr{d}_{N}\otimes \deltau^\mr{d}_{N} \end{bmatrix}\in\R^{\dnu\dnp\times N-1},\hspace{\savespaceeqeight}\\
	\hspace{\savespaceeqeight}X_\Delta   & = \begin{bmatrix} \deltax^\mr{d}_2 & \cdots  & \deltax^\mr{d}_{N} \end{bmatrix}\in\R^{\dnx\times N-1},\\
	\hspace{\savespaceeqeight}\Xp_\Delta & = \begin{bmatrix} p^\mr{d}_2\otimes \deltax^\mr{d}_2 & \cdots &  p^\mr{d}_{N}\otimes \deltax^\mr{d}_{N} \end{bmatrix}\in\R^{\dnx\dnp\times N-1}, \hspace{\savespaceeqeight}\\
	\hspace{\savespaceeqeight}\Xsdelta   & = \begin{bmatrix} \deltax^\mr{d}_3 & \cdots  & \deltax^\mr{d}_{N+1} \end{bmatrix}\in\R^{\dnx\times N-1},
\end{align}
\end{subequations}
where `$\kron$' denotes the Kronecker product. \new{Moreover, for $\mG:=\begin{bmatrix} X_\Delta^\top & {\Xp_\Delta}^\top & U_\Delta^\top & {\Up_\Delta}^\top \end{bmatrix}^\top$, define $\ddataset{N}$ being \emph{persistently exciting} (PE) if $\mG$ has full row-rank, i.e., $\mr{rank}(\mG) = (1+\dnp)(\dnx+\dnu)$.}

Consider the  \emph{velocity controller} in terms of the LPV control law}
\begin{equation}\label{eq:fblaw}
    \deltau_k = K^\mr{v}(p_k)\deltax_k = \begin{bmatrix} K_0^\mr{v} & \bar{K}^\mr{v}\end{bmatrix}\begin{bmatrix} \deltax_k \\ p_k\otimes \deltax_k\end{bmatrix},
\end{equation}
with  $K^\mr{v}(p_k)=K^\mr{v}_0 + \sum_{i=1}^{\dnp}K^\mr{v}_i p_{i,k}$ and $\bar{K}^\mr{v}=\begin{bmatrix} K^\mr{v}_1 & \cdots & K^\mr{v}_{\dnp} \end{bmatrix}$. \new{Interconnection of this controller with the embedded velocity-form \eqref{eq:NL_sys_LPV}, can be formulated as a fully data-driven closed-loop representation. This is summarized in the following Corollary, derived from \cite[Thm.~1]{Verhoek2022_DDLPVstatefb}.}
\begin{corollary}\label{cor:closed-loop-data-based-general}
	Given a PE $\mc{D}_N^\Delta$ generated by \eqref{eq:NL_sys} based on which $\Xsdelta$ and $\mG$ are constructed. %
	Then, the interconnection of \eqref{eq:NL_sys_LPV}, i.e., \eqref{eq:NL_sys_velocity}, and {a given} $K^\mr{v}(p_k)$ under the feedback law \eqref{eq:fblaw} is represented equivalently as
	\begin{equation}\label{e:data-based-CLLPV-state-feedback-general}
		\deltax_{k+1} = \Xsdelta \mc{V} \begin{bmatrix} \deltax_k \\ p_k\otimes \deltax_k \\ p_k\otimes p_k\otimes \deltax_k \end{bmatrix},
	\end{equation} 
	where $\mc{V}\in\mathbb{R}^{N-1 \times \dnx(1+\dnp+\dnp^2) }$ is any matrix that satisfies
	\begin{equation}\label{e:consist-cond-general}
		\begin{bmatrix}
    		I_{\dnx} & 0& 0\\
    		0& I_{\dnp}\otimes I_{\dnx} & 0\\
    		K^\mr{v}_0 & \bar{K}^\mr{v} & 0\\
    		0 & I_{\dnp}\otimes K^\mr{v}_0 &  I_{\dnp}\otimes \bar{K}^\mr{v}
		\end{bmatrix} = \mG \mc{V}.
	\end{equation}
\end{corollary}
\begin{proof}
    The proof follows directly from \cite[Thm~1]{Verhoek2022_DDLPVstatefb}.
\end{proof}
We can now apply the direct data-driven LPV state-feedback controller synthesis methods from \cite{Verhoek2022_DDLPVstatefb} to \new{\emph{synthesize} the LPV velocity controller $K^\mr{v}(p_k)$}
for the system \eqref{eq:NL_sys_LPV}, i.e., a controller for the velocity-form of \eqref{eq:NL_sys}.

\vspace{\rmwhitebfsec}\section{Data-driven state-feedback control of NL systems with guarantees}\label{sec:mainresults}\vspace{\rmwhiteafsec}
\vspace{-0.1mm}\subsection{Data-driven velocity state-feedback synthesis}\vspace{-0.1mm}
Using the closed-loop data-driven representation of the velocity-form of \eqref{eq:NL_sys}, %
{we now formulate %
synthesis of $K^\mr{v}(p_k)$ with the objectives of  %
stabilization of \eqref{eq:NL_sys_velocity} and optimal performance in terms of the quadratic infinite-time horizon cost}
\begin{equation}\label{eq:infhorcost}
    J(\deltax,\deltau) = {\textstyle\sum_{k=1}^\infty} \deltax_k^\top Q \deltax_k + \deltau_k^\top R \deltau_k,
\end{equation}
where $Q,R\posdef0$ are user-defined matrices that encode the performance expectations. \chris{\emph{Velocity-dissipativity} of the closed-loop system (the velocity-form \eqref{eq:NL_sys_velocity} driven by the feedback law \eqref{eq:fblaw}) w.r.t. the supply function $\meu{S}_\mr{v}(\deltau_k, \deltax_k) = -(\deltax_k^\top Q \deltax_k + \deltau_k^\top R \deltau_k)$ implies that \eqref{eq:infhorcost} is finite. The following Corollary derived from \cite[Thm.~4]{Verhoek2022_DDLPVstatefb} gives a fully data-based algorithm for the synthesis of a velocity controller $K^\mr{v}(p_k)$ that ensures this and even minimizes \eqref{eq:infhorcost}.}
\begin{corollary}\label{cor:synthesis}
    Given a PE $\ddataset{N}$ generated by \eqref{eq:NL_sys}.
    \chris{Let $Z=Z^\top\in\mathbb{R}^{n_\mathrm{x}\times n_\mathrm{x}}$, with $Z\posdef0$, be the minimizer of $\sup_{p\in\mb{P}}\mr{trace}(Z)$, {such that there exist} multipliers $\mc{F}\in\mb{R}^{N-1\times\dnx(1+\dnp+\dnp^2)}$, $F_Q\in\mb{R}^{(N-1)(1+\dnp)\times \dnx (1+\dnp)}$, $\Xi\in\R^{4\dnx\dnp\times4\dnx\dnp}$, $Y_0\in\R^{\dnu\times\dnx}$, {and} $\bar{Y}\in\R^{\dnu\times\dnx\dnp}$}
    \begin{subequations}\label{eq:synthesis_conditions}
    satisfying
    \allowdisplaybreaks
	\begin{gather}
		\left[\begin{array}{c} * \\ \hline  * \end{array}\right]^\top \left[\begin{array}{c|c} \Xi & 0 \\ \hline  0 & W \end{array}\right]^\top	 \left[\begin{array}{c c} L_{11} & L_{12} \\ I & 0 \\ \hline  L_{21} & L_{22} \end{array}\right]\prec 0, \\
		\left[\begin{array}{c} * \\ \hline * \end{array}\right]^\top \underbrace{\begin{bsmallmatrix} \Xi_{11} & \Xi_{12} \\ \Xi_{12}^\top & \Xi_{22} \end{bsmallmatrix}}_{\Xi} \left[\begin{array}{c} I \\ \hline \mc{P} \end{array}\right]   \preceq 0, \quad \Xi_{22} \succ 0,\\
		\begin{bmatrix} Z & 0 & 0 \\ 0 & I_{n_\mr{p}}\otimes Z & 0 \\ Y_0 & \bar{Y} & 0 \\ 0 & I_{n_\mr{p}}\otimes Y_0 &  I_{n_\mr{p}}\otimes \bar{Y} \end{bmatrix} = \mG \mc{F}, \\
		\chris{\mc{F} \begin{bsmallmatrix} I_{\dnx}\\ p\kron I_{\dnx} \\ p\kron p\kron I_{\dnx} \end{bsmallmatrix} = \begin{bsmallmatrix} I_{N-1} \\ p\otimes I_{N-1} \end{bsmallmatrix}^{\!\top}\!\! F_Q \begin{bsmallmatrix} I_{n_\mr{x}} \\  p\otimes I_{n_\mr{x}} \end{bsmallmatrix},} \label{eq:syncond:equalFFQ}
	\end{gather}
    \end{subequations}
    for all $p\in\mb{P}$,
	\new{where %
	$\mc{P} = \diag(p)\otimes I_{2n_\mr{x}}$, and
    \begingroup\allowdisplaybreaks
    \begin{align}
        W & =\begin{bmatrix} Z_0 & F_Q^\top \overrightarrow{\mc{X}}_{\!\Delta}^\top  & \begin{bmatrix}  Q^{\frac{1}{2}}Z  & 0\end{bmatrix}^\top & \mc{Y}^\top R^{\frac{1}{2}} \\ \overrightarrow{\mc{X}}_{\!\Delta} F_Q & Z_0 & 0 & 0 \\ \begin{bmatrix}  Q^{\frac{1}{2}}Z  & 0\end{bmatrix} & 0 & I_{n_\mr{x}} & 0 \\ R^{\frac{1}{2}}\mc{Y} & 0 & 0 &  I_{n_\mr{u}} \end{bmatrix}, \notag\\
        Z_0 & = \mr{blkdiag}(Z, \ 0_{\dnx\dnp\times\dnx\dnp}),\quad\mc{Y} =[\,Y_0 \ \ \bar{Y}\,],  \notag\\
        \overrightarrow{\mc{X}}_{\!\Delta} & = \mr{blkdiag}(\Xsdelta, \ I_{\dnp}\kron\Xsdelta), \notag\\
        L_{11} & = 0_{2n_\mr{xp}\times 2n_\mr{xp}}, \hspace{3.88mm}\ L_{12}  =  \begin{bmatrix} 1_{\dnp}\otimes I_{2\dnx} & 0_{2n_\mr{xp}\times n_\mr{xu}} \end{bmatrix}, \notag\\
	    L_{21} & = \begin{bmatrix} 
	              0_{\dnx\times 2n_\mr{xp}}\\ 
	              I_{\dnp}\otimes \Gamma_1\\ 
	              0_{\dnx\times 2n_\mr{xp}}\\ 
	              I_{\dnp} \otimes \Gamma_2\\ 
	              0_{n_\mr{xu}\times 2n_\mr{xp}}
	           \end{bmatrix}, \ L_{22}  =%
	           \begin{bmatrix}
	              \Gamma_1 & 0 \\
	              1_{\dnp}\otimes 0_{\dnx\times 2n_\mr{x}} & 0\\
	              \Gamma_2 & 0 \\
	              1_{\dnp}\otimes 0_{\dnx\times 2n_\mr{x}} & 0 \\
	              0 & I_{n_\mr{xu}}
	           \end{bmatrix}, \notag\\ 
	    \Gamma_1 & = \begin{bmatrix} I_{\dnx} & 0 \end{bmatrix}, \hspace{8.9mm}\Gamma_2 = \begin{bmatrix} 0 & I_{\dnx} \end{bmatrix}, \label{eq:LFT-LQR}
    \end{align}%
    \endgroup%
    with $n_\mr{xp}=n_\mr{x}n_\mr{p}$, $n_\mr{xu}=n_\mr{x}+n_\mr{u}$.}
	Then, the state-feedback controller $K^\mr{v}(p_k)$ with $K^\mr{v}_0 = Y_0 Z^{-1}$, and $\bar{K}^\mr{v} = \bar{Y} (I_{n_\mr{p}}\otimes {Z} )^{-1}$ is a stabilizing controller for \eqref{eq:NL_sys_LPV}, and achieves the minimum of \eqref{eq:infhorcost} over all initial conditions $\deltax_1\in\mathbb{R}^{n_\mathrm{x}}$ and scheduling trajectories $p\in\mathbb{P}^{\mathbb{N}}$.
\end{corollary}
\begin{proof}
    See \cite[Thm.~4]{Verhoek2022_DDLPVstatefb}.
\end{proof}
\chris{Note that \eqref{eq:syncond:equalFFQ} can be easily satisfied by defining $\mc{F}=[\mc{F}_1 \,\mc{F}_2\,\mc{F}_3]$ in terms of a permutation of $F_Q=\begin{bsmallmatrix} F_{11} & F_{12} \\ F_{21} & F_{22} \end{bsmallmatrix}$, where $\mc{F}_1 = F_{11}$, $\mc{F}_2$ is constructed from the rows and columns of $F_{21}$ and $F_{12}$, respectively, and $\mc{F}_3$ is a permutation of $F_{22}$.}
By reformulation of \eqref{eq:synthesis_conditions} and assuming that $\mb{P}$ is compact, the synthesis algorithm of Corollary~\ref{cor:synthesis} corresponds to a \emph{semi-definite program} (SDP) with a finite set of \emph{linear matrix inequality} (LMI) constraints. The resulting %
controller $K^\mr{v}(p_k)$ provides stability and performance guarantees for the LPV surrogate form under all possible variations of $p$. This --through the embedding principle-- implies stability and performance in terms of Definition~\ref{def:velocity_stab},~\ref{def:velocity_diss} of the closed-loop velocity-form \eqref{eq:NL_sys_velocity} with {$K^\mr{v}(\psi(x_k, u_{k}, x_{k-1}, u_{k-1}))$} where $p_k$ {is substituted by} \eqref{eq:computation_p}. Hence, using \emph{only} the data-dictionary $\mc{D}_N^\Delta$ from the NL system \eqref{eq:NL_sys}, we synthesized a NL controller for the velocity-form, which corresponds to our contribution~C1. The problem that remains is to show that there exists a NL controller $K^\textsc{nl}$ for which $K^\mr{v}$ is its velocity-form, enabling to prove that that applying $K^\textsc{nl}$ on the unknown system \eqref{eq:NL_sys} will imply USAS and USD guarantees of the closed-loop operation.

\vspace{\rmwhitebfssec}\subsection{Realization of the NL controller}
\begin{figure}
    \centering
    \vspace{2mm}
    \includegraphics[width=0.8\linewidth]{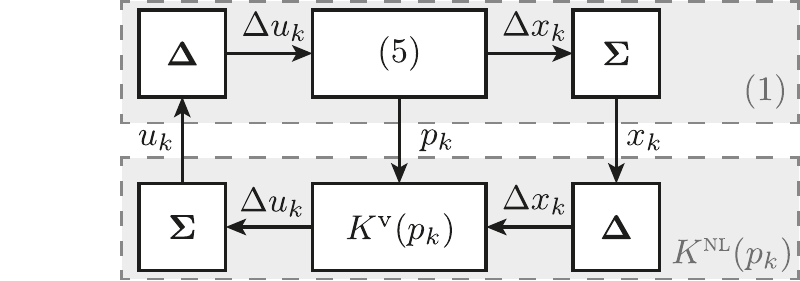}
    \vspace{-0mm}
    \vspace{-2mm}
    \caption{Realization of the controller.}\label{fig:realization}
    \vspace{-7mm}
\end{figure}

\vspace{\rmwhiteafssec}
{For the controller realization, we use the time-difference and summing operators $\difop$ and $\sumop$ on signals, such that $\sumop\deltax_k=x_k$, $\difop x_k = \deltax_k$ and $\difop(\sumop\deltax_k)=\deltax_k$. Note that these are the DT equivalents of the time-integration and differentiation operators in \emph{continuous-time} (CT). Hence, if we apply these to the closed-loop as depicted in Fig.~\ref{fig:realization}, we can define the NL controller {as}} %
\begin{equation}\label{eq:primalcontroller}
    \hspace{-1mm}K^\textsc{NL}:\left\{\begin{aligned}
        \chi_{k+1} & = \begin{bmatrix} 0 & 0 \\ - K^\mr{v}(p_k)  & I \end{bmatrix} \chi_{k} + \begin{bmatrix} I \\ K^\mr{v}(p_k) \end{bmatrix} x_k, \\
        u_k & = \begin{bmatrix} - K^\mr{v}(p_k)  & I \end{bmatrix} \hspace{0.8mm}\chi_{k} + \hspace{1.8mm} K^\mr{v}(p_k) \hspace{2mm}x_k, \\
        p_k & = \psi(x_k, u_{k}, \chi_{k}),
    \end{aligned}\right.
\end{equation}
where $\chi_k = \begin{bmatrix} x_{k-1}^\top & u_{k-1}^\top \end{bmatrix}^\top$. {This is easily derived by noting that $u_k = \deltau_k+u_{k-1}$, i.e., }
\begin{subequations}\label{eq:controlrealization}
\allowdisplaybreaks
\begin{align}
    \deltau_k & = K^\mr{v}(p_k)\deltax_k, \\
    (u_{k}-u_{k-1}) & = K^\mr{v}(p_k) (x_{k}-x_{k-1}), \\
    u_{k} & = K^\mr{v}(p_k) (x_{k}-x_{k-1}) + u_{k-1}.\label{eq:controlrealization:c}
\end{align}
\end{subequations}
{Hence, the interconnection of $K^\mr{v}(p_k)$ with \eqref{eq:NL_sys_velocity} is in fact the velocity-form of the interconnection of \eqref{eq:primalcontroller} with \eqref{eq:NL_sys}.}
Note that to compute $u_k$ in the output equation of $K^\textsc{NL}$, $p_k$ is also dependent on $u_k$. This means that computation of $u_k$ requires the solution of a \emph{fixed point problem}, for which many reliable solvers exist, or one can use
$u_{k-1}$ instead of $u_k$ in the computation of $p_k$ as an approximative solution.

\vspace{\rmwhitebfssec}\subsection{Stability and performance guarantees}\label{ss:stabandperf}\vspace{\rmwhiteafssec}
With the realization of the controller for the original form of the NL system established, we are ready to present the main {result of the paper:} %
\begin{theorem}\label{thm:inducedstab}
    Given a PE $\mc{D}_{N}^{\Delta}$ from \eqref{eq:NL_sys}, based on which a stabilizing controller $K^\mr{v}$ is synthesized via Corollary~\ref{cor:synthesis}. Then, the interconnection of the realized controller $K^\textsc{nl}$ \eqref{eq:primalcontroller}  and the NL system \eqref{eq:NL_sys} is guaranteed to be USAS.
\end{theorem}
\begin{proof}
    With the synthesis of $K^\mr{v}$, we know that the velocity-form \eqref{eq:NL_sys_velocity} in closed-loop with $K^\mr{v}$ is asymptotically stable. {Realization of the controller $K^\textsc{nl}$ ensures that  its velocity-form is $K^\mr{v}$ and \eqref{eq:NL_sys} in closed-loop with $K^\textsc{nl}$ has a velocity-form that is the interconnection of \eqref{eq:NL_sys_velocity} with $K^\mr{v}$. Under these conditions, asymptotic stability of the velocity-interconnection implies USAS of the closed-loop interconnection of \eqref{eq:NL_sys} with $K^\textsc{nl}$ based on \cite[Thm.~8.3]{Koelewijn2023}.}
\end{proof}
\begin{conjecture}
    Given a PE $\mc{D}_{N}^{\Delta}$ from \eqref{eq:NL_sys} based on which a stabilizing controller $K^\mr{v}$ is synthesized via Corollary~\ref{cor:synthesis}. Then, the interconnection of the realized controller $K^\textsc{nl}$ \eqref{eq:primalcontroller} and the NL system \eqref{eq:NL_sys} is USD w.r.t. the supply function $\meu{S}_\mr{s}(u_k, u_\ast, x_k, x_\ast) = -(x_k-x_\ast)^\top Q (x_k-x_\ast) - (u_k-u_\ast)^\top R (u_k-u_\ast)$.
\end{conjecture}
We introduced the implication of performance as a conjecture, because the link between velocity-dissipativity and general USD has not been formally proven -- only under certain technical conditions, see \cite[Sec.~8.3]{Koelewijn2023}. {However, the analysis of USD through the velocity-form shares strong similarities with analysis of a stronger dissipativity notion called \emph{incremental dissipativity} \cite{Koelewijn2023}. %
Hence, there are strong indications that velocity-dissipativity w.r.t. a quadratic supply function implies USD w.r.t. a quadratic supply function}.

Finally, we want to note that the universal shifted controller {guarantees convergence} to an equilibrium point $(x_\ast,u_\ast)\in\ms{E}$. To ensure that the system is driven to a desired equilibrium point, {we can add integrators to {$K^\mr{v}$}, see \cite[Cor.~8.2]{Koelewijn2023} for the description of the approach and the example in Section \ref{sec:examples}}.

\vspace{\rmwhitebfsec}\section{Simulation study}\label{sec:examples}\vspace{\rmwhiteafsec}
We demonstrate the applicability of our results on a simulator of an unbalanced disc system, for which we synthesize a universal shifted data-driven state-feedback controller and compare it with a data-driven state-feedback LPV controller that uses a direct LPV embedding of the NL system, \new{cf.~\cite{Verhoek2022_DDLPVstatefb}}. For {comparison,} %
we also synthesize an LTI {data-driven} controller. %
The CT dynamics of the unbalanced disc system mimic those of an inverted pendulum and are thus described by the following ordinary differential equation
\begin{equation}\label{eq:unbalanced-disc}
    \ddot{\theta}(t)=-\tfrac{mgl}{J}\sin(\theta(t))-\tfrac{1}{\tau}\dot{\theta}(t)+\tfrac{K_\mr{m}}{\tau}u(t),
\end{equation}
where $\theta$ is the angular position of the disc in radians, $u$ is the input voltage to the system, which is its control input, and $m,g,l,J,\tau,K_\mr{m}$ are the physical parameters of the system that we take from \cite[Tab.~I]{Verhoek2022_DDLPVstatefb}. Discretizing the dynamics using a first-order Euler method and writing {them} in the form of \eqref{eq:NL_sys} gives
\begin{subequations}\label{eq:DT-unbaldisc}
\begin{align}
    \hspace{-1mm}x_{1,k+1} & = x_{1,k} + T_\mr{s} x_{2,k},\\
    \hspace{-1mm}x_{2,k+1} & = (\tfrac{T_\mr{s}}{\tau}-1)x_{2,k} -\tfrac{T_\mr{s}mgl}{J}\sin(x_{1,k})+\tfrac{T_\mr{s}K_\mr{m}}{\tau}u_k,
\end{align}
where $x_k = \begin{bsmallmatrix} \theta_k & \dot\theta_k \end{bsmallmatrix}^\top$.
\end{subequations}
We choose the sampling-time as $T_\mr{s}=0.01$ [s], which gives a negligible discretization error through the Euler scheme. The control objective is to design a controller that tracks a reference for $\theta_k$ with zero steady-state error, which requires integrator action. We introduce the integrator behavior with the tuning parameter $\alpha$, see~\cite[Cor.~8.2]{Koelewijn2023}. For the direct LPV design, we introduce integrator behavior by adding an augmented state $x_{\mr{aug},k+1}=\alpha x_{\mr{aug},k} + \theta_{\mr{ref},k} - x_{1,k}$. Note that with the extra state, we require a larger data-dictionary for the construction of the direct data-driven LPV representation.

\begin{figure*}
\deflen{removewhitespaceforfig}{-5mm}
	\centering
	\begin{minipage}[t]{0.49\linewidth}
	   \vspace{2mm}
	   \includegraphics[scale=1, trim=0mm 0.95mm 0mm 0.7mm, clip]{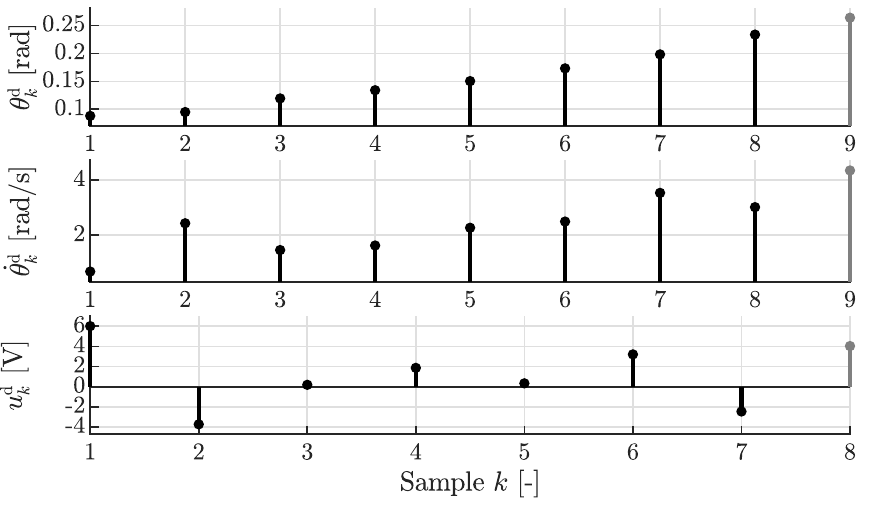}
	   \vspace{\removewhitespaceforfig}
	   \vspace{-2mm}
	   \caption{Data-dictionary $\mc{D}_{N\!+\!1}^{\textsc{nl}}$ used for the NL and LPV control synthesis {with} $N=8$. The {extra} gray (\legendline{cgray}) {data-points} are required  for the direct LPV representation because of the added state for the {integrator} behavior.}
	   \label{fig:datadictionary}
	\end{minipage}\hfill
	\begin{minipage}[t]{0.49 \linewidth}
       \vspace{2mm}
	   \includegraphics[scale=1, trim=0mm 0.95mm 0mm 0.7mm, clip]{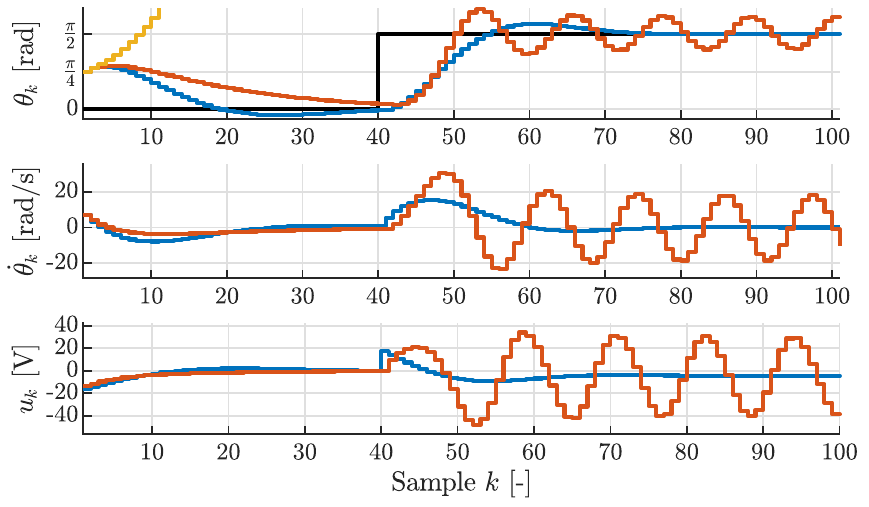}
	   \vspace{\removewhitespaceforfig}
	   \vspace{-2mm}
	   \caption{Response of the unbalanced disc with the universal shifted controller (\legendline{mblue}) and the LPV controller (\legendline{morange}) in closed-loop for a step reference (\legendline{black}). An LTI controller (\legendline{myellow}) designed with the same specifications diverges directly.}%
	   \label{fig:resexample}
	\end{minipage}\vspace{\removewhitespaceforfig}
\end{figure*}

The velocity-form of \eqref{eq:DT-unbaldisc}
can be computed analytically:
\begin{subequations}\label{eq:DT-unbaldisc:velocity}
\begin{align}
    \deltax_{k+1} & = A_\mr{v}(x_{1,k}, x_{1,k-1})\deltax_k + B_\mr{v}\deltau_k, \\ 
    \deltay_k & = \deltax_k,
\end{align}
\end{subequations}
where $B_\mr{v}= \begin{bmatrix} 0 & \tfrac{T_\mr{s}K_\mr{m}}{\tau} \end{bmatrix}^\top$ and 
\begin{equation*}
    A_\mr{v}(x_{1,k}, x_{1,k-1}) = \begin{bmatrix} 1 & T_\mr{s} \\ -\tfrac{T_\mr{s}mgl}{J}\sind(x_{1,k}, x_{1,k-1}) & 1-\tfrac{T_\mr{s}}{\tau} \end{bmatrix},
\end{equation*}
{with} $\sind(a,b):=\tfrac{\sin(a)-\sin(b)}{a-b}$\new{, which is obtained by solving the integral in \eqref{eq:NL_sys_velocity}}. For the data-driven design of the NL universal shifted controller, we choose\footnote{{This basis is used} for simplicity and comparison purposes with the direct LPV design, but one could alternatively choose a polynomial basis.} $p_k := \psi(x_{1,k}, x_{1,k-1})=\sind(x_{1,k}, x_{1,k-1})$, which allows for an LPV embedding of the velocity-form \eqref{eq:DT-unbaldisc:velocity}. Note that $\lim_{x_{1,k}\rightleftarrows x_{1,k-1}}\psi(x_{1,k}, x_{1,k-1})$ exists and for all {trajectories of \eqref{eq:DT-unbaldisc}} $\psi(x_{1,k}, x_{1,k-1})\in[-1, 1]$. Hence, we take this interval as $\mb{P}$. For the {direct data-driven} LPV design, we follow \cite{Verhoek2022_DDLPVstatefb} {to formulate an LPV embedding of \eqref{eq:DT-unbaldisc}} where we choose $p_k = \tfrac{\sin(x_{1,k})}{x_{1,k}}$, {which is well-defined for $x_{1,k}=0$}. 

We are now ready to construct the LPV data-driven representations and synthesize controllers for the velocity-form and the {original} system. To construct well-posed data-driven representations for both approaches, while using the same data-set, we need $\mr{rank}(\mc{G}) = (1+\dnp)(\dnx+1+\dnu)=8$, i.e., we need $N\ge8$. The data-dictionary $\mc{D}_{N+1}^{\textsc{nl}}$ is obtained by applying {white noise} $u^\mathrm{d}_k\sim\mc{N}(0,3)$ to \eqref{eq:DT-unbaldisc} under an initial condition $x^\mathrm{d}_1\sim\mc{U}(0,1)$. {The resulting $\mc{D}_{N+1}^{\textsc{nl}}$} is shown in Fig.~\ref{fig:datadictionary}, where the additional data-points required for the augmented LPV representation are {given in} gray.
{Using} $\mc{D}_{N+1}^{\textsc{nl}}$, we construct the direct data-driven LPV representation as in~\cite{Verhoek2022_DDLPVstatefb} and for the velocity-form we construct \eqref{eq:datamatrices} and verify that indeed $\mr{rank}(\mG) = (1+\dnp)(\dnx+\dnu)=6$, %
{giving a well-posed} data-driven representation of \eqref{eq:DT-unbaldisc:velocity}. 

Using the constructed representations, we design an LPV controller and a universal shifted controller (with integral action) using \cite[Thm.~4]{Verhoek2022_DDLPVstatefb} and Corollary~\ref{cor:synthesis}, respectively, with the tuning parameters $Q=I$, $R=2$, $\alpha=0.9$. 
Running the synthesis algorithms yield \new{the parameters of} %
$K^\mr{v}$ and $K^\textsc{lpv}$.
We want to highlight here that  \eqref{eq:DT-unbaldisc} in closed-loop with  the universal shifted controller with $K^\mr{v}$ as above implies USAS of the closed-loop system, %
while the LPV controller only guarantees stability %
of the origin of the NL closed-loop system. {The} latter {is} problematic {for} reference tracking \cite{Koelewijn2020_pitfalls}, which we showcase in the following simulation study.

We simulate\footnote{See \texttt{youtu.be/NeOC9PBipMY} for an animation of the simulations.} \eqref{eq:DT-unbaldisc} in closed-loop with the LPV and universal shifted controller for the initial condition $x_{1}=\begin{bmatrix} \tfrac{\pi}{4} & 5 \end{bmatrix}^\top$. The system must follow a step-reference of magnitude $\tfrac{\pi}{2}$, which pushes the closed-loop away from the origin. The simulated responses of the closed-loops are plotted in Fig.~\ref{fig:resexample}, which shows that both controllers can regulate the system back to the origin. However, when the step reference is applied, only the universal shifted controller can drive the system to the reference, while the LPV controller ends up in a limit cycle. \new{We also design a data-driven LTI state-feedback controller using \cite[Thm.~4]{dePersisTesi2020}} under the same performance specifications and data. \new{Note that the LTI data-driven design spans an LTI behavior based on $\mc{D}_{N+1}^{\textsc{nl}}$, which results in a local approximation of the NL system. %
As outside of this local range, the LTI behavior is not valid anymore, the stability guarantee fails and the closed-loop system quickly diverges with the LTI controller}, see
Fig.~\ref{fig:resexample}.

This example shows that we can synthesize state-feedback controllers for general NL systems of the form \eqref{eq:NL_sys} that are \new{universally shifted} stabilizing and performing while using \emph{only} measured data from the system and a given a set of basis functions $\psi$ that is assumed to span the nonlinearities.

\vspace{\rmwhitebfsec}\section{Conclusions}\label{sec:conclusion}\vspace{\rmwhiteafsec}
By connecting {results} on velocity-dissipativity/stability and universal shifted dissipativity/stability with data-driven {controller design}, we have shown that the data-driven velocity-form of a general NL system with full state-observation enables direct data-driven control of NL systems with \new{equilibrium independent} stability and performance guarantees. The elegance {and effectiveness} of this concept is demonstrated on a simulation example of {a NL unbalanced disc} system. The presented concepts in this paper can be seen as the first approach that achieves direct data-driven analysis and control in the general NL setting. For future research, we aim to use the data-driven velocity-form for general NL input-output-representations and derive the corresponding analysis and synthesis methods {under a dynamic output-feedback setting}. \new{Moreover, handling noise and the correct choice for $\psi$ are interesting open problems.}

\bibliographystyle{IEEEtran}
\bibliography{ref_cdc2023_1}

\end{document}